\documentclass[10pt, conference, letterpaper]{IEEEtran}

\usepackage{booktabs}

\usepackage{cite}
\usepackage{url}
\usepackage{graphicx}
\usepackage{xcolor}
\usepackage{subfigure}

\usepackage{diagbox} 
\usepackage{tabularx}

\usepackage[colorlinks,citecolor=blue]{hyperref}

\usepackage{float}
\usepackage{multirow}

\usepackage{url}

\usepackage{breakurl}

\begin{document}
\title{A machine learning approach for detecting CNAME cloaking-based tracking on the Web}

\author{
    \IEEEauthorblockN{Ha Dao}
    \IEEEauthorblockA{The Graduate University for Advanced Studies 
    (Sokendai)\\
    Email: hadao@nii.ac.jp}
    \and
    \IEEEauthorblockN{Kensuke Fukuda}
    \IEEEauthorblockA{NII/Sokendai\\
    Email: kensuke@nii.ac.jp}
}

\maketitle

\begin{abstract}
Various 
in-browser privacy protection techniques have been designed to protect
end-users from third-party tracking.
In an arms race against these counter-measures, the tracking providers 
developed a new technique called CNAME cloaking based tracking to avoid issues 
with browsers that block 
third-party cookies and requests.
To detect this tracking technique, browser
extensions
require on-demand DNS lookup APIs. 
This feature is however only supported by the Firefox browser.

In this paper, we propose a supervised machine learning-based method to detect 
CNAME cloaking-based tracking without the on-demand DNS lookup.
Our goal is to detect both sites and requests linked to CNAME cloaking-related 
tracking.
We crawl a list of target sites and store all HTTP/HTTPS requests with their 
attributes. 
Then we label all instances automatically by looking up CNAME record of
subdomain, and applying wildcard matching based on well-known tracking filter 
lists. 
After extracting features, we build a supervised classification model to 
distinguish site  and  request related to CNAME cloaking-based tracking.
Our evaluation shows that the proposed approach outperforms well-known tracking 
filter lists: F1 scores of 0.790 for sites and 0.885 for requests.
By analyzing the feature permutation importance, we demonstrate that the number of 
scripts and the proportion of XMLHttpRequests are discriminative for 
detecting sites, and the length of URL request is helpful in detecting requests.
Finally, we analyze concept drift by using the 2018 dataset to train a model 
and obtain a reasonable performance on the 2020 dataset for detecting both 
sites and requests using CNAME cloaking-based tracking. 
\end{abstract}

\section{Introduction}
There are several reasons websites intend to track user's browsing activities.
In some acceptable cases, it is simply to make the user browsing experience 
faster and more convenient.
On the other hand, user's browsing activities are also used to determine user 
browsing habits and preferences for advertisement and analytics 
on the web, which can be frightful for privacy-sensitive users 
\cite{mayer2012third}.

Aiming to capture the economic value of user browsing habit through tracking, 
many entities 
design services to collect user's information, 
even without any user's knowledge or consent \cite{acar2014web} \cite{ 
yen2012host} \cite{metwalley2015online}. 
There are some effective approaches against third-party tracking for user's 
privacy-preserving.
Ad-blocking and third-party tracking protections have been successfully 
preventing such trackers using blacklist \cite{Ghostery} 
\cite{DisconnectExtension} \cite{ublock}, machine learning-based
detection \cite{wu2015trackerdetector}
\cite{metwalley2015unsupervised}, and others 
\cite{kushmerick1999learning} \cite{roesner2012detecting}
\cite{pan2015not}.

However, third-party web tracking has been getting more sophisticated in an 
arms race against these counter-measures. 
One of the current trends is the exploitation of Canonical Name Record or Alias 
(CNAME) record in DNS, to bypass the filter lists in browsers and 
extensions. 
This technique, called \emph{CNAME cloaking-based tracking}, uses CNAME to 
disguise requests to a third-party tracker as first-party ones.
The current counter-measures mainly focus on third-party
tracking and
do not impact 
first-party resources, 
CNAME cloaking-based tracking can thus
bypass 
this blocking.

There are some existing methods to detect CNAME cloaking-based
tracking.
EasyPrivacy\cite{privacy}, AdGuard tracking protection\cite{AdGuard}, and other 
filter lists manually add new first party subdomains which are fronts for CNAME 
cloaking to these blacklists.
However, this approach will dramatically increase the size of the blacklists 
and these subdomains need to be updated frequently.
Besides that, uBlock Origin since version 1.24.1b0 performs a DNS lookup of the 
hostname loading a resource to determine if the underlying subdomain is related 
to CNAME cloaking or not.
Nevertheless, only Firefox allows uBlock Origin to block CNAME cloaking 
because the other browsers do not support DNS resolution API \cite{ublock}.

In this paper, we propose a machine learning approach to detect CNAME
cloaking-based tracking on the web without dynamic DNS lookup.
Our method consists of three steps.
(1) We crawl a list of target sites and store all HTTP requests with
their attributes. 
Then we look up CNAME record of subdomains, and label all
instances by applying wildcard matching based on well-known tracking filter 
lists on this CNAME records (\autoref{sec:data_collection}).
(2) Next, we extract their features (\autoref{sec:feature_extraction}). 
(3) Finally, we build a supervised
classification model to distinguish sites and requests using CNAME
cloaking-based tracking (\autoref{sec:model}).

The contributions of this study are summarized as follows:
\begin{itemize}
    \item We propose an effective method to detect CNAME cloaking-based    
    tracking. 
    We obtain a F1 score of 0.790 for detecting CNAME cloaking sites 
    and of 0.885 for requests (\autoref{sec:general_result}).
    \item We introduce site-related and request-related
    features for CNAME 
    cloaking detection (\autoref{sec:feature_extraction}).
    We also make a detailed analysis about the discriminative power of features 
    regarding CNAME cloaking-based tracking detection 
    (\autoref{sec:feature_permutation}).
    \item We show that our supervised model also works against newly crawled 
    data by analyzing concept drift (\autoref{sec:concept_drift}).
\end{itemize}

\section{Background and related work}
\subsection{Background}
\subsubsection{Browser-based privacy protection techniques}
Several browser-based privacy protections have been designed to protect 
end-users from third-party tracking, such as Ghostery\cite{Ghostery}, 
Disconnect\cite{DisconnectExtension}, and uBlock Origin\cite{ublock}.
They work effectively to detect third-party tracking.

\subsubsection{CNAME cloaking-based tracking}
\begin{figure}[ht!]
    \centering
    \includegraphics[width=0.45\textwidth]{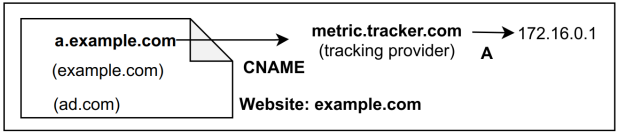}
    \vspace{-0.2cm}
    \caption{Overview of CNAME cloaking-based tracking.}
    \label{fig:definition}
\end{figure}
The usage of DNS CNAME records coupled with Content Delivery Network is 
increasingly commonplace to improve site load times, reduce bandwidth costs, 
and increase content availability and redundancy.
CNAME has also been used for user tracking. 
Tracking providers ask their clients to delegate a subdomain for data 
collection and tracking, and link it to an external server using a CNAME 
DNS record \cite{adobeDocument}.
It behaves differently from ordinary third-party tracking because it uses the 
first-party subdomains; browsers and extensions generally do not impact 
these resources.

For example, when end-users access a site \emph{example.com}, this site embeds 
a first-party tracker \emph{a.example.com}, which points to a tracking provider 
\emph{tracker.com} via the CNAME \emph{metric.tracker.com}.
Tracking provider \emph{tracker.com} thus track activities of end-users on the 
site \emph{example.com} (\autoref{fig:definition}).

\subsection{Term definitions}
\subsubsection{Request linked to CNAME cloaking-related
tracking}
An HTTP request by a subdomain is used for collecting, and sharing information 
about a particular user's activities on the web by using CNAME 
cloaking-based tracking.

\subsubsection{Site with CNAME cloaking-related tracking occurring}
A site sets up a subdomain that points to a tracking provider domain using a 
CNAME record. 

\subsection{Related work}
\subsubsection{Anti third-party tracking techniques}
The privacy hazards of online web tracking have been studied extensively.

\emph{Machine learning-based tracking detection:} Harass et 
al.~\cite{metwalley2015unsupervised} develop an unsupervised detection method 
that inspects URL queries in HTTP(S) requests to detect tracking activities.
Yamada et al.~\cite{yamada2010web} analyze traffic at the network gateway to 
monitor all tracking sites in the administrative network and constructs a graph
between sites and their visited time to detect tracking sites.
Wu et al.~\cite{wu2015trackerdetector} develop DMTrackerDetector which 
automatically detects third-party trackers offline to efficiently generate
blacklists using structural hole theory and supervised machine learning. 

\emph{Non-machine learning-based tracking detection}: Schelter and 
Kunegis~\cite{schelter2016tracking} perform a large-scale analysis of 
third-party trackers by extracting third-party embeddings from more than 41 
million domains to study global online tracking.
Roesner et al.~\cite{roesner2012detecting} develop a client-side method for 
detecting and classifying five kinds of third-party trackers on the 500 most popular and 500 less popular sites according to the 
Alexa ranking.
To cut off the tracking chain of third-party web tracking, Pan et al. 
~\cite{pan2015not} develop TrackingFree which isolates unique identifiers into
different browser principals so that the identifiers still exist but are not 
unique among different websites.

\subsubsection{Existing in-browser CNAME cloaking-based tracking detection techniques}
In order to block first-party trackers using CNAME records, uBlock Origin 
resolves the hostname of a DNS record to determine if an underlying subdomain 
is related to CNAME cloaking or not.
However, only Firefox with DNS API allows uBlock Origin to block the CNAME 
cloaking \cite{ublock}.
Besides that, the well-know tracking filter lists make an continuous effort to 
manually update first-party subdomains which are fronts for CNAME cloaking to 
these blacklists.
It makes day-to-day filter lists updating tedious and time-consuming.
\section{Design and implementation}
\label{sec:design_model}
Here, we describe  our 
method to detect CNAME cloaking-based tracking.
\subsection{Method overview}
\begin{figure}[ht!]
    \centering
    \vspace{-0.3cm}
    \includegraphics[width=0.45\textwidth]{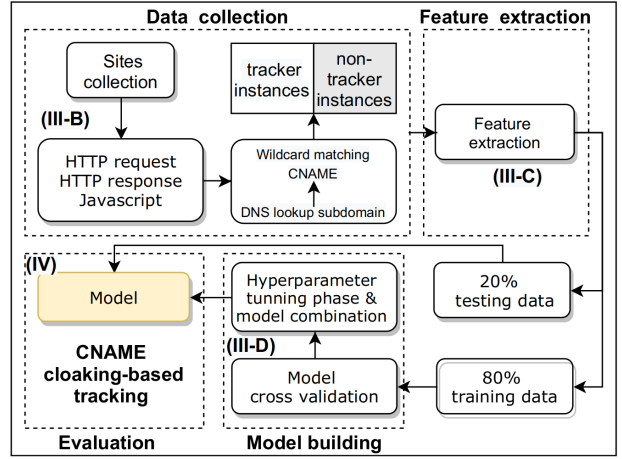}
    \vspace{-0.2cm}
    \caption{Processing flow.}
    \label{fig:overview}
\end{figure}
Our goal is to develop a CNAME cloaking-based tracking detection method using 
machine learning.
Our method flags sites and requests related  to  CNAME cloaking-based 
tracking.

As shown in ~\autoref{fig:overview}, the method consists of four steps: data collection, feature extraction, model development, and evaluation.
\begin{enumerate}
    \item Crawl the target sites, including sites with and without CNAME 
    cloaking based tracking, and save all the HTTP requests and their 
    attributes.
    Then we perform DNS lookup for all subdomains, and label all instances 
    automatically by matching well-known tracking filter lists with CNAME 
    behind all subdomain as described in \cite{dao2020}.
    These instances are divided into two sets, which we call the tracker set 
    and the non-tracker set (\autoref{sec:data_collection}).
    \item Extract features for sites and requests 
    (\autoref{sec:feature_extraction}).
    \item Check the F1 score of 10 classification algorithms using 10-fold 
    stratified cross-validation. 
    After evaluating performance, we select some effective classification 
    algorithms for hyper-parameter tuning phase, and combine conceptually 
    different machine learning classification algorithms to build a model
    (\autoref{sec:model}). 
    \item Evaluate the models with the testing data (\autoref{sec:results}).
\end{enumerate}

\subsection{Data collection}
\label{sec:data_collection}
\subsubsection{Finding CNAME cloaking-based tracking with blacklists}
\label{cname_blacklist_approach}
We present our method to label CNAME cloaking-based tracking. 
Firstly, we perform DNS queries for CNAME answer set on all subdomains of each 
site. 
Then, we apply wildcard matching based on well-known tracking filter lists to 
each resolved CNAME in order to detect CNAME cloaking base tracking.
More specifically, we select \emph{third-party tracking domains} and 
\emph{international third-party tracking domains} from Easy privacy
list\cite{EasyPrivacyList} and 
\emph{tracking servers list} from AdGuard tracking protection 
filter\cite{AdGuard} as of April 26, 2020.
This approach was used in the previous work~\cite{dao2020}.

\subsubsection{Datasets}
We use OpenWPM\cite{Englehardt2016Online} with vanilla Firefox in 
April 2020
to crawl 1,762 sites from Alexa Top 300k where CNAME 
cloaking-based tracking was previously detected \cite{dao2020}.
We also crawl 1,762 sites that were randomly picked from all sites without 
CNAME cloaking-based tracking \cite{dao2020}.
The stored data is all HTTP/HTTPS requests and their responses for each site.
Then, applying the method (\autoref{cname_blacklist_approach}) to the stored 
crawled data, we find that 4,047 HTTP requests in 1,532 sites used CNAME 
cloaking-based tracking\footnote{The dataset is available at \url{https://github.com/fukuda-lab/cname-cloaking}}.

We also rely on another dataset crawled on Alexa Top 100K sites using
OpenWPM\cite{Englehardt2016Online} in April 2018.
We check the historical forward DNS (FDNS) datasets provided by
Rapid7~\cite{Rapid7DNS}, and apply the methodology in 
\autoref{cname_blacklist_approach} to detect CNAME cloaking-based tracking in this historical data.
We find that 2,546 HTTP requests in 1,005 sites are related to CNAME 
cloaking-based tracking.
We also use 1,005 additional randomly picked sites without CNAME-cloaking from 
this dataset to analyze concept drift in \autoref{sec:concept_drift}.
The details of the 2020 dataset and the 2018 dataset are listed in 
\autoref{tab:crawled_data}.

\begin{table}[ht!]
\centering

\caption{Summary of data: 2,010 sites in 2018 and 3,524 sites in 2020.}
\vspace{-0.2cm}
\label{tab:crawled_data}
\scalebox{0.95}{
\begin{tabular}{llrr}
    \toprule
    Metrics            &                    & 2018            & 2020\\
    \midrule
    3rd party requests
                   &                    & 185,896 (62.2\%) & 276,686 (63.8\%)\\
    1st party requests 
                   & domain             &  60,278 (20.2\%) & 120,689 (27.8\%)\\
                   & \textbf{subdomain} &  52,476 (17.6\%) &  36,399  (8.4\%)\\
    \midrule
    \textbf{Total requests} &           & 298,649 (100\%)  & 433,774 (100\%) \\
    \midrule
    \textbf{Total sites}    &           &   2,010          & 3,524 \\
    \bottomrule
\end{tabular}
}
\vspace{-0.3cm}
\end{table}

\subsection{Feature extraction}
\label{sec:feature_extraction}
We experimentally extract the
following 
features related to sites and requests.
\subsubsection{Site with CNAME cloaking-related tracking occurring}
\label{sec:feature_site}

\begin{itemize}
    \item \emph{num\_url}:
    The total number of HTTP requests in a site.
     \item \emph{num\_1st}:
    The total number of  first-party requests in a site.
     \item \emph{num\_3rd}:
    The total number of
    third-party requests in a site.
    \item \emph{pct\_script\_call}: The proportion of fingerprinting-related method calls in a site.
        
    \item \emph{pct\_xhr}: The proportion of XMLHttpRequest in a site, which is 
    a built-in browser object that can be used to make HTTP requests in 
    JavaScript to exchange data between the web browser and the server.
        
    \item \emph{pct\_3rd\_window}: The proportion of requests originating from 
    a third-party DOM window in the window hierarchy \cite{nsIContentPolicy}.
        
    \item \emph{ranking}: 
    The site ranking according to the Alexa ranking \cite{Alexa}.
    \item \emph{country}: 
    The site country 
    based on Free IP Geolocation API\cite{iplocation}.
    \item \emph{category}: 
    The site 
    category according to the FortiGuard Web Filtering \cite{FortiGuard} dataset in  
    April 2020.
 
\end{itemize}

\subsubsection{Request linked to CNAME cloaking-related tracking}
\begin{itemize}
    \item \emph{method}:
    The desired action to be performed for given request.
    \item \emph{is\_xhr}:
    The request uses an API that provides scripted client functionality for 
    transferring data between a client and a server. 
    \item \emph{content\_type}: The HTML tag that resulted in a request, 
    such as image, javascript or document, which are defined in this IDL 
    \cite{nsIContentPolicy}.
    \item \emph{len\_url}:
    The length of request URL.
     \item \emph{len\_sub}:
    The subdomain length of request.
    \item \emph{len\_prefix\_sub}:
    The subdomain prefix length of request.
    \item \emph{num\_prefix\_sub}:
    The number of subdomain prefixes.
    \item \emph{prefix\_sub\_blacklist}:
    The subdomain prefix is among subdomain prefixes in tracking filter lists 
    \cite{AdGuard}\cite{EasyPrivacyList}.
    \item \emph{is\_sub\_dic}:
    The prefix of subdomain is a word in English dictionary.
    \item \emph{entropy\_url}:
    The randomness of
    request URL by calculating the metric entropy from request URL.
    \item \emph{entropy\_sub}:
    The randomness of
    subdomain by calculating the metric entropy from subdomain.
    \item \emph{entropy\_prefix\_sub}:
    The randomness of
    subdomain prefix by calculating the metric entropy from subdomain prefix.
\end{itemize}

\subsection{Modeling and preliminary results}
\label{sec:model}
First of all, we split the 2020 dataset (\autoref{tab:crawled_data}) into 
testing data and training data.
The percentage of the data held over for testing is 20\%.
It is used in ~\autoref{sec:general_result} to evaluate our model.
Next, we describe how to build a classification model to detect CNAME 
cloaking-based tracking using testing data (80\% of the 2020 dataset).

\subsubsection{Model cross-validation}
\label{model_cross_validation}
We randomly split the training dataset into 10 smaller sets (folds) without 
replacement, where nine folds are used for the model training and the remaining 
one fold is for validating. 
After obtaining 10 performance estimates by repeating this procedure ten 
times, we take their average as the final performance estimate.
We then compare 10 classification algorithms and evaluate their F1 score using 
this stratified cross-validation procedure on the training data (80\% the 
dataset 2020).
We employ 10 supervised classification algorithms with scikit-learn
module (random\_state = 2 with default parameters):
\begin{itemize}
    \item Support Vector Classification (SVC)
    \item Decision Tree
    \item AdaBoost
    \item Random Forest
    \item Extra Trees
    \item Gradient Boosting
    \item Multi-layer Perceptron (MLP)
    \item K-Nearest Neighbors (KNN)
    \item Logistic Regression
    \item Linear Discriminant Analysis (LDA)
\end{itemize}

\begin{figure*}
  \centering
  \subfigure[Site linked to 
    CNAME cloaking-based tracking.]
    {\label{fig:cross_site_score}\includegraphics[width=0.45\textwidth]{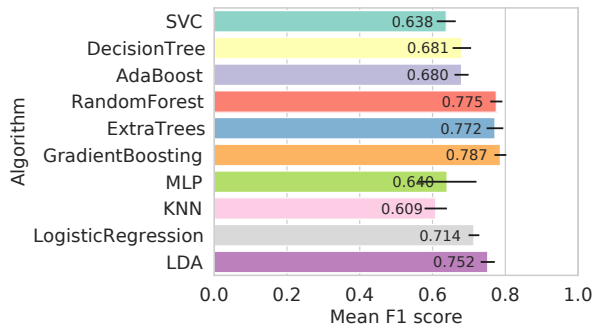}}\qquad
  \subfigure[Request linked to 
    CNAME cloaking-based tracking.]
    {\label{fig:cross_request_score}\includegraphics[width=0.45\textwidth]{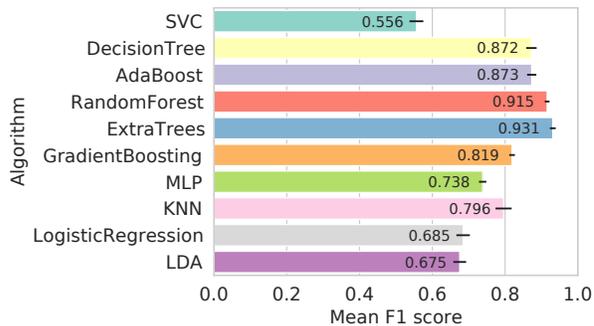}}
\caption{The F1 score for the 10 selected classification algorithms using 
    10-fold stratified cross-validation in 2020 dataset regarding the detection of 
    CNAME cloaking-based tracking.
    The mean and standard deviation are computed on the 10 folds of the 
    cross-validation.}
    \vspace{-0.3cm}
\end{figure*}
We use the F1 score for evaluating the performance of the classifiers. 
Larger values of the F1 score ($\approx$ 1.0) indicate better performance, and 
lower values ($\approx$ 0) correspond to worse performance. 
\autoref{fig:cross_site_score} shows the F1 scores for the 10 selected 
algorithms using 10-fold stratified cross-validation in the 2020 dataset for 
detecting sites with CNAME cloaking-related tracking occurring. 
All classification algorithms have a different detection performance.
The most effective classification algorithms are Random Forest, Extra 
Trees, and Gradient Boosting, 
while Support Vector Classification, Multi-layer Perceptron, and
K-nearest Neighbor achieve moderate 
performances for this
dataset.

~\autoref{fig:cross_request_score} shows the F1 scores 
for detecting requests linked to CNAME cloaking-related tracking.
The most
effective classification algorithms are Random Forest and 
Extra Trees,
while SVC shows the worst
performance.

\subsubsection{Hyper-parameter tuning for best models}
From the previous performance evaluations, 
we select \emph{RandomForest, ExtraTrees, and GradientBoosting} classifiers 
for the hyper-parameter tuning phase for site classification detection.
For requests, we select \emph{Random Forest and Extra Trees} classifiers.

We perform a grid search optimization for all of them regarding F1 score.
The parameter setting that gave the best results on the training data are shown in
\autoref{tab:hyper-parameter}.

\begin{table}[t!]
    \centering	
    \caption{Best parameter settings in training phase regarding F1 score.
    }	
    \vspace{-0.2cm}
    \label{tab:hyper-parameter}	
    \begin{tabular}{llrrr}	
        \toprule	
        \bf{Algorithm}  & \bf{Parameter}&   \bf{Site} & \bf{Request}\\
        \midrule                                        
        \emph{Random}     & max\_features       & 1&1\\ 
        \emph{Forest}     & min\_samples\_split & 8&2 \\
                   & min\_samples\_leaf  & 3&1\\
                   & n\_estimators       & 100&300 \\
        \midrule    
        Extra      & max\_features       & 10&1 \\
        Trees      & min\_samples\_split & 2&2 \\
                   & min\_samples\_leaf  & 3&1 \\
                   & n\_estimators       & 300&100 \\
        \midrule             
        \emph{Gradient} & max\_features     & 0.3&Not selected\\
        \emph{Boosting} & min\_samples\_leaf    & 100&         \\
                 & n\_estimators         & 300&      \\
                 & learning\_rate        & 0.2&  \\
                 & max\_depth            & 8&             \\
        \bottomrule	
    \end{tabular}	
    \vspace{-0.3cm}
\end{table}

\subsubsection{Model combination}
Now, we have a set of closely well-performing models for 
detecting site and requests linked to CNAME cloaking-based tracking.
In order to balance out their individual weaknesses, we combine conceptually 
different machine learning classification algorithms and and use the sum of predicted
probabilities to predict the class labels.

We thus use VotingClassifier with \emph{soft} voting to combine the predictions 
coming from  the following supervised classification algorithms: RandomForest, 
ExtraTrees, and GradientBoosting for detecting sites, and RandomForest and 
ExtraTrees for detecting requests. 
The performance of separate algorithms and the combined model are
discussed in the next section.

\section{Classification performance analysis}
\label{sec:results}
We evaluate our model and feature importance regarding detection 
of CNAME cloaking-based tracking.
We also investigate concept drift using our datasets from 2018 to 2020.

\subsection{General results}
\label{sec:general_result}
\subsubsection{The performance of model}
The results obtained using the test set for sites and requests 
linked to CNAME cloaking-based 
tracking detection on 20\% of the 2020 dataset (preserved for this test) are
shown in \autoref{tab:Results_score_final_site} and
\autoref{tab:Results_score_final_request}.
We first show that the F1 score of all classification algorithms improve slightly after the
hyper-parameter tuning for both sites and request.
Next, the combined model achieves 0.790 of F1 score for detecting sites and 0.885 
of F1 score for detecting requests.
These results show that the combination model exhibit similar performance 
to other classifiers regarding detection of sites linked to 
CNAME cloaking based-tracking, but that combined model works better 
for detecting requests. 
In order to ease following result comparison, we use the combined models for further analyses.

Checking false negatives, we find that 
some sites/requests linked to CNAME cloaking have the same
attributes with sites/requests without CNAME cloaking based tracking. 
For example, the \emph{healthpartners.com} linked to 
tracking provider Adobe, with only six scripts and the proportion
XMLHttpRequest in this site is 0.07.
For request, the request \emph{https://gjr5.yoigo.com/ea.js} points to
tracking provider \emph{Eulerian}; its prefix \emph{gjr5} is not in the blacklists and it does not contain any user identification. 
However, we believe that if we crawl more data with the request linked
to track provider \emph{Eulerian}, we can classify 
this request as CNAME cloaking-based tracking.

\begin{table}[ht!]
    \vspace{-0.3cm}
    \centering
    \caption{Results obtained using the test set for site linked to CNAME cloaking-based 
    tracking detection.
    }
    \vspace{-0.1cm}
    \scalebox{0.85}{
    \begin{tabular}{lrrrr}
        \toprule
        \bf{Score}  & \bf{RandomForest}&\bf{ExtraTree}  & \bf{GradientBoosting} & \bf{Combined} \\
                    & Default/Tuned     & Default/Tuned & Default/Tuned  & \\
        \midrule
        Precision & 0.777/0.784   &     0.771/0.763    &0.781/\emph{0.801}&0.794    \\ 
        Recall    & 0.730/0.748  &     0.736/0.770 &0.779/\emph{0.782}&0.785  \\
        F1 score  & 0.753/0.766    &  0.753/0.766    &0.780/\emph{0.791} &0.790\\
        \bottomrule
    \end{tabular}
    }
    \label{tab:Results_score_final_site}
\end{table}

\begin{table}[ht!]
    \vspace{-0.3cm}
    \centering
    \caption{Results obtained using the test set for request linked to CNAME cloaking-based 
    tracking detection.
    }
    \vspace{-0.1cm}
    \begin{tabular}{lrrr}
        \toprule
        \bf{Score}  & \bf{Random Forest}&\bf{ExtraTree}  & \bf{Combined} \\
                    & Default/Tuned     & Default/Tuned  & \\
        \midrule
        Precision &  0.940/0.935  &     0.928/0.949     &\emph{0.949}    \\ 
        Recall    &0.807/0.830    &     0.811/0.802     &\emph{0.828} \\
        F1 score  &0.869/0.879    &  0.866/0.870        &\emph{0.885}\\
        \bottomrule
    \end{tabular}
    \label{tab:Results_score_final_request}
\end{table}

\subsubsection{Comparison with the well-known tracking filter lists}
To block CNAME cloaking-based tracking without DNS resolution, well-known 
tracking filter lists such as the EasyPrivacy list and the AdGuard tracking 
filter list include the 
first-party subdomains which are
fronts for CNAME cloaking. 
We compare the request 
detection performance between our machine learning approach and these tracking 
filter lists in \autoref{tab:comparison}. 
We confirm that our method outperforms the evaluated blocking lists.
Meanwhile, almost all browsers (except Firefox) do not provide a DNS resolution 
API which is critical CNAME cloaking-based tracking detection\cite{dao2020}. 
Beyond automating blacklist building, our approach is thus especially helpful 
to block CNAME cloaking-based tracking.

\begin{table}[ht!]
    \vspace{-0.3cm}
    \centering
    \caption{A comparison of request detection performance.}
    \vspace{-0.2cm}
    \begin{tabular}{lrrrrr}
        \toprule
        \bf{Method}         & \bf{Precision} & \bf{Recall} & \bf{F1 score}  \\
        \midrule
        Machine learning        & \bf{0.949} &  \bf{0.828} &    \bf{0.885} \\
        Easy Privacy list       & 0.880      &       0.721 &         0.792 \\
        Adguard tracking filter &	0.918      &       0.417 &         0.574 \\
        \bottomrule
    \end{tabular}
    \label{tab:comparison}
    \vspace{-0.3cm}
\end{table}

\subsection{Feature permutation importance}
\label{sec:feature_permutation}
We investigate 
the permutation importance\cite{breiman2001random} to discuss the feature importance of the selected classifier.
\begin{figure*}
  \centering
  \subfigure[Site linked to 
    CNAME cloaking-based tracking.]
    {\label{fig:feature_site}\includegraphics[width=0.46\textwidth]{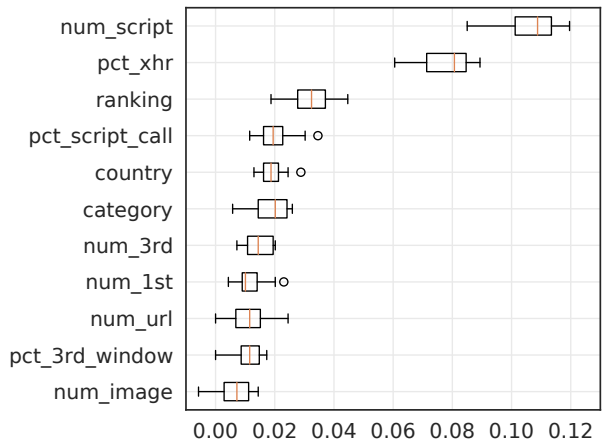}}\qquad
  \subfigure[Request linked to 
    CNAME cloaking-based tracking.]
    {\label{fig:feature_request}\includegraphics[width=0.46\textwidth]{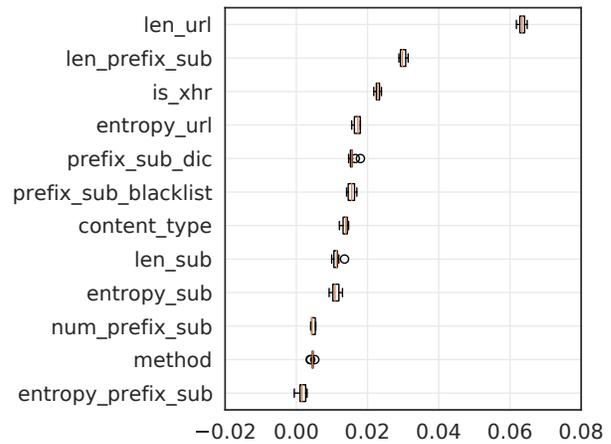}}

    \caption{Permutation importance of the combined model for CNAME cloaking-related tracking occurring.
    The box extends from the lower to upper quartile values of the data, 
    with a line at the median.
    The number of times a feature is randomly shuffled is n\_repeats = 10.}
    \vspace{-0.5em}

\end{figure*}

\autoref{fig:feature_site} represents the feature permutation importance 
regarding site detection.
We observe that the number of script (\emph{num\_script}) and the proportion of 
XMLHttpRequest requests (\emph{pct\_xhr}) are the most useful features.
In addition, tracking providers are spread across 
different types of sites \cite{dao2020},
site category (\emph{category}) and site country (\emph{country}) are thus not helpful in detecting sites with CNAME 
cloaking-related tracking occurring. 
 
\autoref{fig:feature_request} shows the feature permutation importance of the 
model for detecting the requests. 
The result reveals that the length of request URL (\emph{len\_url}) has the highest importance.
We assume that almost requests with a subdomain
used for CNAME cloaking-based tracking have a
bigger length than the request used for collecting
content of site, because they contain user’
identification.
Also, the subdomain prefix length (\emph{len\_prefix\_sub}), the XHR usage 
request (\emph{is\_xhr}), and the randomness of URL request (\emph{entropy\_url}) are discriminative features for request detection.
In addition to that, the subdomain prefix blacklist presence (\emph{prefix\_sub\_blacklist}) is not effective 
to detect requests.
This is because some publishers also use the same subdomain prefix that is used for 
both CNAME cloaking and other non-tracking resources.

\subsection{Concept drift analysis}
\label{sec:concept_drift}
Finally, we investigate the ability to detect sites  and  requests related  to  
CNAME cloaking-based tracking in the latest dataset (2020) using a model 
trained on the old dataset (2018). 
We use the 2018 dataset to train a model and test it on new data collected in 
2020 (\autoref{tab:crawled_data}).
We apply the method explained in \autoref{sec:model} to build a model
and obtain the 
results in \autoref{tab:results_2018/2020}.
\begin{table}[ht!]
    \centering
    \vspace{-0.3cm}
    \caption{Results obtained using the new site/request in 2020 dataset to 
    evaluate 2018 dataset model.}
    \vspace{-0.1cm}
    \begin{tabular}{lrrr}
        \toprule
        \bf{Instance type} & \bf{Precision} & \bf{Recall} & \bf{F1 score} \\
        \midrule
        Site        &        0.608 &    0.787   &  0.686  \\
        Request     &        0.935    & 0.558  &   0.699 \\
        \bottomrule
    \end{tabular}
    \label{tab:results_2018/2020}
\end{table}

Our result shows that the performance degradation is limited. 
Specifically, F1 score for sites with CNAME cloaking-related 
tracking decreases by 0.104, and by 0.186 for requests.

To explain this degradation, 
we examine the 2018 and 2020 datasets.
In 2018, we do not see many 
random subdomain prefixes and short requests, while it is the case in 2020.
For sites, we observe a decreasing number of requests, 
including third and first-party request for each site in the 2020 
dataset.
These changes can be plausible reasons for the degradation in our 
model.
Besides that, with rapid changes of web technology, tracking providers 
might also adjust their target site and change the implementation methods to deploy CNAME 
cloaking-based tracking. 
Although the performance degradation is limited between 2018 and 2020, 
periodic model retraining can alleviate this problem if more detection accuracy is required.

\section{Conclusion and discussion}
Recently, CNAME cloaking-based tracking on the web has attracted
much attention.
The current best countermeasure, blacklist approach,
strongly dependant 
on realtime name resolution.
In this paper, we proposed a machine learning approach to detect sites and HTTP 
requests containing CNAME cloaking-based tracking. 
Through the comprehensive analysis, we demonstrated the effectiveness of our 
method. 

The proposed method complements existing techniques for CNAME cloaking-based 
tracking detection and it can help researchers, browser extensions 
developers, and blacklist maintainers 
to build highly 
effective systems against this tracking technique.
Once tracking providers know the existence of the
proposed detection methodology, they could attempt to evade it. 
In future work, we intend to 
add new features 
to improve our detection performance. 

\section*{Acknowledgement}
We thank Johan Mazel for his valuable comments.

\bibliographystyle{IEEEtran}
\bibliography{main.bbl}

\newpage
\end{document}